\documentclass[amssymb,longbibliography,aps,prb]{revtex4-2}
\usepackage{amsmath}
\usepackage{mathrsfs}
\usepackage{graphicx}
\usepackage{xcolor}

\begin{document}
\title{Light Induced Orbital Magnetism in Metals via Inverse Faraday Effect }
\author{Priya Sharma}
\affiliation{Department of Physics and Institute for Materials Science, University of Connecticut, Storrs, CT 06269, USA}
\affiliation{Advanced Technology Institute and Department of Physics, University of Surrey, Guildford GU1 7XH, UK}
\author{Alexander V. Balatsky}
\affiliation{Department of Physics and Institute for Materials Science, University of Connecticut, Storrs, CT 06269, USA}
 \affiliation{Nordita, KTH Royal Institute of Technology and Stockholm University, Hannes Alfv\'{e}ns v\"{a}g 12, SE-106 91 Stockholm, Sweden}

\begin{abstract}
We present a microscopic calculation of the inverse Faraday effect in metals. We derive a static local magnetic moment  induced on the application of high-frequency light, using the Eilenberger formulation of quasiclassical theory. We include the effect of disorder and formulate a theory applicable across the entire temperature range, in the absence of external applied fields. For light-induced electric fields of amplitude $\sim 100 kV/cm$, the induced fields  are large, $\sim 0.1 T$ for metallic Nb! The predictions of our theory agree with recent experimental and theoretical results \cite{NatPhotonics:2020,Oppeneer:2018}. An extension of this approach to superconductors would open a new route of inducing orbital magnetic field and potentially vortices in superconductors.
\end{abstract}

\maketitle

The ultrafast manipulation of magnetic states is of technological importance in non-volatile magnetic storage, ubiquitous in modern day device applications~\cite{Metals:2019}. Light-induced magnetism~\cite{FerromagMoire:2022}, light-induced superconductivity~\cite{SCCuprate:2011}, light-induced translational and hidden orders~\cite{CDWNatPhys:2020} have been investigated as examples of optical manipulation of quantum states. The demands on pushing up the read and write speeds in magnetic storage media are ever-increasing. The state-of-the-art NVMe technology boasts up to subnanosecond read/write speeds ($\sim 7.5 GB/sec$ for Gen 4 NVMe drives) with intrinsic limitations set by the generation of magnetic fields by current. The search for technologies to control magnetization without the use of magnetic fields is accelerating with optical means emerging as a promising leader~\cite{RoadMap:2022}. The Nobel Prize in Physics 2023 highlights the emergence of attosecond pump-probe methods to study electron dynamics in quantum matter. With the advent of THz laser technology,  the ability to manipulate magnetic order on ultrafast sub-picosecond timescales is being explored for magnetic switching/reversal ~\cite{AdvQuantTech:2022,Sc.Reps:2017}. The physics that governs light-matter interactions on these ultrafast timescales makes available a route to quantum states inaccessible in equilibrium, that can be technologically transformative.

Pitaevskii~\cite{Pitaevskii:1960} first proposed the effect of a time-varying electric field on the stress tensor in a dispersive medium. Based on thermodynamic arguments, he deduced that the dielectric tensor in a homogenous medium has non-zero off-diagonal components for circularly(or elliptically, for that matter) polarized light i.e., for time-dependent electric fields with rotating polarization. This represents the induction of an effective induced magnetization by light with a rotating polarization. ~\cite{Malmstrom:1965} first reported an observed induced magnetization in various liquids and glasses via the  "Inverse Faraday Effect"(IFE), coining the name. This suggested the interesting approach of transduction viz., transferring the  orbital momentum of light to the electronic degrees of freedom in matter. Further, it offers a means for {\it{rectification}} of quantum order to generate quantum features that did not exist in the absence of dynamical fields. Such an effect has since been studied extensively. An effective magnetization, as suggested by Pitaevskii, could be induced by a number of disparate microscopic schemes. The majority of schemes explored thus far have involved induced magnetic fields by orientation of the spin degree of freedom in a wide range of systems. In such cases, the angular momentum of the incident circularly polarized light is transferred to the underlying system by orientation of system spins i.e., as an induced spin magnetization.  Historically, the understanding of magnetization dynamics involving spin degrees of freedom included spin-orbit(SO) interactions. The SO interaction sets the time-scale for this dynamics. The transfer of angular momentum from light to the underlying spins in the material is realised through various SO mechanisms : Rashba spin-orbit coupling \cite{MetalsTHzT0:2011}; through magneto-electric effects that simulate such a coupling for example, via an axial magnetoelectric effect \cite{AME:2021}; through crystal symmetry breaking for example, in polar crystals \cite{PhysRevMaterials.1.014401,Edelstein:1998}; through optical transitions leading to transfer of spin between Zeeman split spin-orbit  bands among others \cite{SOsplitZeemantransfer:2016}. In these schemes, the spins of electrons respond via SO coupling to an effective magnetization originating from electric currents set up by the driving field. These effects can be potentially large and a number of these schemes have been explored for applications such as magnetization switching.

In this paper, we evaluate a spin-independent and purely orbital method to instantiate and further manipulate magnetic  order in metals dynamically by interaction with THz light via the IFE.   This non-dissipative magneto-optical effect is known to photoinduce  static magnetic moments in various materials for circularly polarized light.
 The induced magnetic field corresponds to the response of electrons to the rotating time-dependent electric field of the driving light giving circulating electric currents that represent the static induced magnetic moments. This can be realised in the absence of any  spin-orbit coupling and is completely agnostic to the spin degree of freedom of electrons in the metal. 

\begin{figure}
    \centering
    \includegraphics[width=0.3\textwidth]{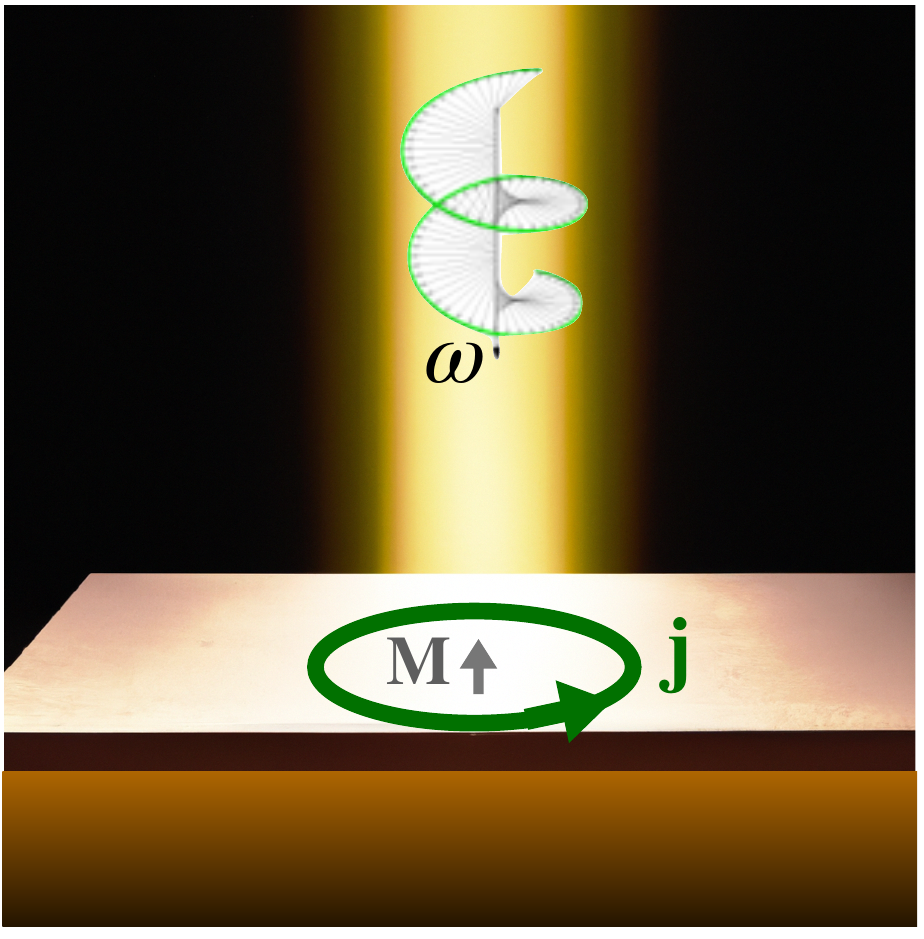}
    \caption{Schematic of geometry to observe induced magnetization in metals via the IFE. Circularly polarized light is shone on the metal surface. (Image generated with help from OpenAI’s DALL-E tool)}
    \label{fig:Schematic}
\end{figure}


  The IFE has been theoretically discussed ~\cite{Malmstrom:1966, Popova:2011, Popova:2012} and  observed experimentally in garnets and magnetic materials where spin magnetization is induced on the application of high-frequency light ~\cite{RevModPhys:2010}. {\it{Ab initio}} methods are the leading theoretical techniques to investigate optically induced spin magnetization dyamics~\cite{RevModPhys:2010}. A semiclassical theory for the orbital IFE in metals was derived by Hertel ~\cite{Hertel:2006}  which has been further developed recently ~\cite{Majedi:2021} and applied to superconductors, where the creation of vortices by light has been proposed phenomenologically. Buzdin  and coworkers suggest vortex generation via the Kibble-Zurek mechanism by quenching into the superconducting state~\cite{Buzdin:2022}. Ginzburg Landau(GL) theory has been used to calculate the induced orbital moments in superconductors and proposed for the manipulation of vortices using the IFE ~\cite{BuzdinAVortices:2022,Yokoyama:2018}. Floquet theory has been applied to study the orbital IFE in Mott insulators ~\cite{Banerjee2022}.

In this paper, we present a microscopic calculation of the orbital IFE applicable to metals and extensible to superconductors. We use quantum many-body theory to address the extent to which we can controllably produce orbital magnetization in a metal using circularly polarized light. We find the effect to be significant and the induced moments predicted by our theory are large. Our microscopic theory offers better degree of control via material parameters such as the plasma frequency of the material $\omega_p$; it is applicable over the entire temperature range, unlike phenomenological and GL theories. It applies on quasiclassical length scales that are microscopic on the scale of the spot size of incident light, but long compared to atomic length scales. In our knowledge, the orbital IFE has not been formulated in terms of the Green's functions for electrons in metals thus far. We propose this comprehensive approach for light manipulation of quantum materials, emphasizing the  potential emergence of novel quantum ordered states and topological excitations via a light-specific mechanism.

We consider a metal exposed to circularly polarized electromagnetic radiation of frequency $\omega$, as shown in Fig.\ref{fig:Schematic}. The primary response of electrons(quasiparticles) is to the oscillating electric field given by $\bf{E} = \bf{E}_0 e^{{{i\bf{K}.\bf{R}}} - i\omega t}$. We find the induced magnetization density via the IFE is given by,
\begin{equation}
\label{M-ind}
{\bf{M}}_{ind} = \mu_B \frac{\omega_p^2}{2\omega(\omega^2 + 16\Gamma^2)}\, (i\epsilon_0\,{\bf{E}}_0\times{\bf{E}}_0^{\star})\zeta^2(\frac{\omega}{2k_B T})\,\,\,,
\end{equation}
where $\Gamma$ is the rate at which quasiparticles scatter from impurities/disorder and $\epsilon_0$ and $\mu_B$ are the free-space permittivity and the Bohr magneton, respectively. The function $\zeta^2(x) \leq 1$ over the temperature range and is discussed in Section \ref{Results}.  ${\bf{M}}_{ind}$ is maximal in the clean limit ($\Gamma \rightarrow 0$) and at low temperatures when $\zeta$  is maximal. For high-frequency THz light, the induced magnetization density is given by equation(\ref{M-ind}). For typical metals  we find the effective magnetic field to be on the scale of $0.001 - 0.1$T, and the specific material estimates are given in Table I.   dc  magnetization induction, discussed here, epitomises a rectified quantum order where the dc order is induced via a nonlinear effect upon application of the ac external field.

\section{Theoretical Framework}
\label{Theory}

We consider high-frequency circularly polarized light of frequency, $\omega$ shining on a metal.
The electrodynamic response has been considered by ~\cite{Majedi:2021}, where the effect is described by the hydrodynamics of the electron density given by the London equation.
Here we  consider the corrections to the Green's functions for electrons in response to the light-induced electric field. To simplify the analysis, we neglect the electron spin, the crystal structure (and band structure) of electrons and consider an isotropic Fermi surface. The crystal point group dependent tensors would need to be used for other symmetries. 

We employ the Keldysh formulation ~\cite{Keldysh:1965} of quasiclassical theory ~\cite{Eilenberger:1968, LarkinOvchinnikov:1969} which
describes phenomena that occur on characteristic length scales much larger
than the Fermi wavelength, $k_F^{-1}$ and characteristic time scales much longer
than the inverse Fermi energy $\varepsilon_F^{-1}$ (Here,  $\hbar  = 1$). It is established as a robust formulation for the theoretical description of degenerate fermionic systems and superconductors, in particular. We consider the response of electrons to the oscillating electric field of the electromagnetic wave, $\bf{E} = \bf{E}_0 e^{i\bf{K}.\bf{R} - i\omega t}$. The driving term corresponding to the response to the oscillating magnetic field of the electromagnetic wave is much smaller, viz., $\propto v_F/c \ll 1$, $v_F$ being the Fermi velocity, and we ignore it.  We use the  Eilenberger equations to formulate and derive the second-order current response via the IFE.

The quasiclassical Green's function, $\hat{g}$ is the propagator for quasiparticles with effective mass $m^{\star}$, energy $\varepsilon$ and Fermi momentum ${\bf{p}}_F = m^{\star} {\bf{v}}_F$, given by solutions to the Eilenberger equation~\cite{Eilenberger:1968}. Observables such as the quasiparticle density are calculated from the Keldysh components of the Green's function, $\hat{g}^K$, viz., $n = \int \frac{d\varepsilon}{2\pi} \frac{d^3 p}{(2\pi\hbar)^3} \mathscr{T} (\hat{g}^K)$; here, $\mathscr{T}$ refers to a trace over Nambu and spin indices. We expand $\hat{g}^K$  in the external field $\bf{E}$, $\hat{g}^K = \hat{g}_0^K + \hat{g}_1^K + \hat{g}_2^K$, with $\hat{g}_i^K \propto \mathcal{O}(\bf{E}^i)$ being the $i$-th order correction to $\hat{g}^K$ and evaluate the current response. We are interested in the second-order $\mathcal{O}(\bf{E}^2)$ corrections to the current density, $\bf{j}_2$, averaged over the period of the oscillating field. We show (details in Appendix) that  
\begin{equation}
\label{jIFE}
    {\bf{j}}_2 = e \, \langle \,\bar{n}_1({\bf{p}})\,\, \bar{\bf{v}}_1({\bf{p}}) \,\rangle_{{\bf{{p}}},(2\pi/\omega)}\,\,\,,
\end{equation}
where $\langle ...\rangle_{{\bf{p}},(2\pi/\omega)} \equiv (\frac{\omega}{2\pi})\int dt \int \frac{d^3 p}{(2\pi)^3} $ refers to an average over the Fermi surface and an average over a time-period of the field; $n_1$ and $\bf{v}_1$ are the first-order corrections to the quasiparticle density and velocity, respectively viz., 
\begin{equation}
\label{bar-defn}
    \bar{n}_1({\bf{p}}) \equiv \int {d\varepsilon} \mathscr{T} \hat{g}_1^K({\bf{p}},\varepsilon) \,\,\,\,;\,\,\,\,
    \bar{\bf{v}}({\bf{p}})  \equiv   {\bf{v}}_F\int {d\varepsilon'} \mathscr{T}\hat{\tau}_3 \hat{g}_1^K({\bf{p}},\varepsilon') \,\,\,,
\end{equation}
where $\hat{\tau}$ are Pauli matrices in Nambu space.
We calculate the first-order corrections $\hat{g}_1^K$ from the Eilenberger equations and obtain the time-averaged $\bf{j}_2$, where we average over a two-dimensional isotropic Fermi surface (details in Appendix). $\bf{j}_2$ represents a static current response to an applied oscillating field via the IFE. It contains curl-less terms and a term that is $\propto \nabla \times \bf{M}$. We identify the latter as a magnetization density induced by the IFE. The irrotational terms describe the so-called ponderomotive forces on electrons. 

Equation(\ref{jIFE}) holds for electrons with a three-dimensional isotropic Fermi surface. As discussed in the Appendix,  equation(\ref{j2}) is the complete expression for the time-averaged corrections to the current density  that are of second order in the field, ${\bf{j}}_2 \propto \mathcal{O}({\bf{E}}^2)$ and includes second-order corrections to the quasiparticle density $n_2$ and velocity  ${\bf{v}}_2$, given by $\hat{g}_2^K$. The three-dimensional Fermi surface averages conspire to leave non-zero contributions only from $n_1$ and ${\bf{v}}_1$. This allows us to compute a nonlinear term ${\bf{j}}_2$, using  mere linear response theory. This convenience is, however, not accessible in lower dimensional systems, 2DEGs for example. We report our results for the three-dimensional isotropic case. The effects of lower spatial dimensions and crystal symmetries are not obvious and need to be addressed separately.

\section{Results and Discussion}
\label{Results}

\begin{figure}
    \centering
    \includegraphics[width=0.6\textwidth]{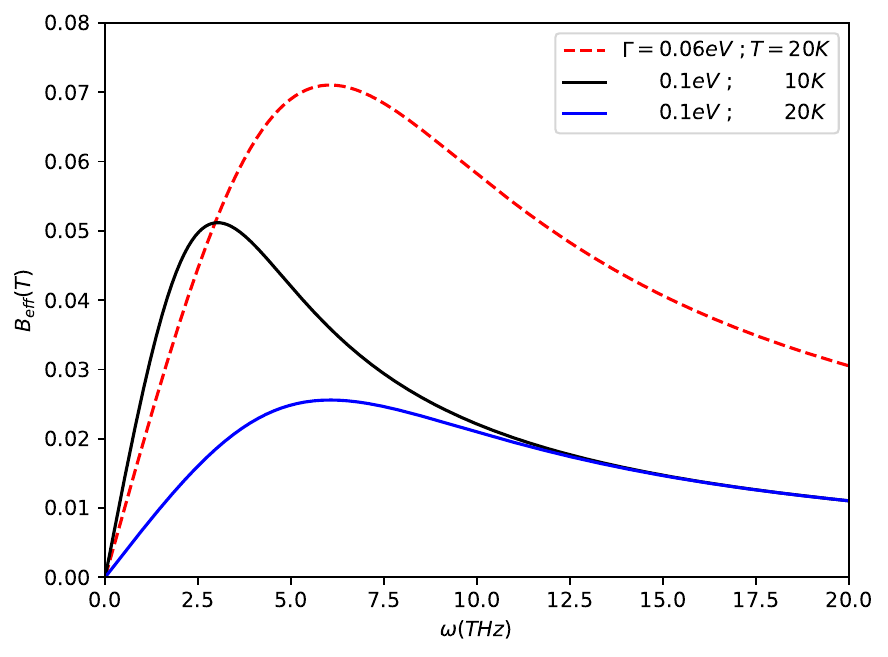}
    \includegraphics[width=0.6\textwidth]{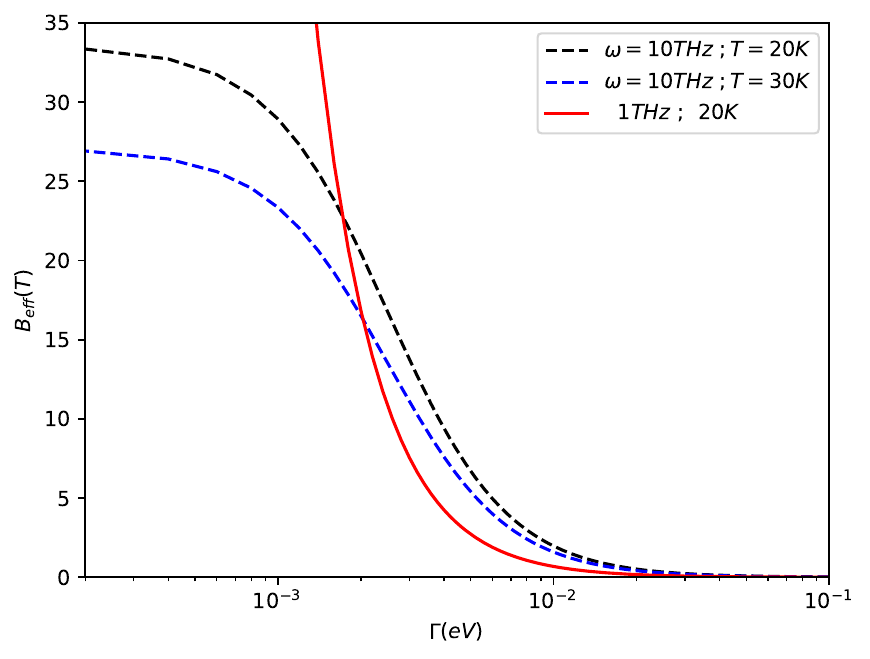}
    \includegraphics[width=0.6\textwidth]{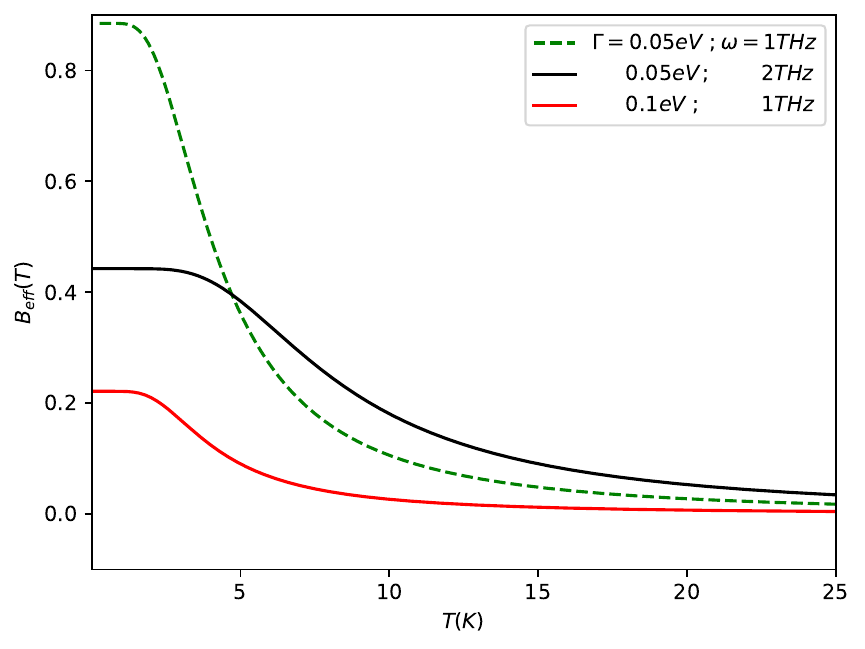}
    \caption{Effective induced magnetic fields $\bf{B}_{eff}$ in units of Tesla for elemental Nb. Plasma frequency $\omega_p = 20 eV$ for Nb and applied electric fields of $E = 100 kV/cm$~\cite{Bonetti:2022} are used here. For reference, $\Gamma = 0.01eV$ corresponds to impurity mean free paths $\sim 0.1 \mu m$.
    } 
    \label{fig:B-eff}
\end{figure}

The induced magnetization density is given by the expression(\ref{M-ind}) with $ \zeta(\beta\omega) = \frac{tanh(\beta\omega)}{\sqrt{2 - tanh^2(\beta\omega)}}$, where $\beta=(2k_B T)^{-1}$. For $\beta\omega \gtrsim 1$, $\zeta \sim 1$ and 
\begin{equation}
    \beta\omega \gtrsim 1\,\,:\,\,{\bf{M}}_{ind} \,\,\rightarrow\,\, \mu_B \frac{\omega_p^2}{2\omega(\omega^2 + 16\Gamma^2)}\, {(i\epsilon_0\,{\bf{E}}\times{\bf{E}}^\star)}\,\,\,.
\end{equation}
In the clean limit($\Gamma=0$), this  agrees with the semiclassical expression derived by Hertel\cite{Hertel:2006}. This also agrees with the classical expression derived by \cite{YOSHINO20112531} in the zero temperature dissipation-less limit.
For $\beta\omega \ll 1$, $\zeta \sim \beta\omega$ and
\begin{equation}
    \beta\omega \ll 1\,\,:\,\,{\bf{M}}_{ind} \,\,\rightarrow\,\, \mu_B \frac{\omega_p^2}{2(\omega^2 + 16\Gamma^2)}\, {(i\epsilon_0\,{\bf{E}}\times{\bf{E}}^\star)}\frac{\omega}{4 (k_B T)^2}
\end{equation}
Therefore, for fixed disorder at a given temperature, the induced magnetization goes through a peak as the frequency $\omega$ is varied. For weak disorder 
set by $\beta\Gamma \lesssim 1$,
\begin{eqnarray}
\label{Born}
    {\bf{M}}_{ind} = \mu_B \frac{\omega_p^2}{2(\omega^2 + 16\Gamma^2)}\, {(i\epsilon_0\,{\bf{E}}\times{\bf{E}}^\star)}\frac{\omega}{4 (k_B T)^2} &\propto& \omega \,\,\,\,\,\,\, \,\,\,;\,\,\,\,\,\beta\omega < \beta\Gamma \ll 1\\
    \nonumber
    = \mu_B \frac{\omega_p^2}{2(\omega^2 + 16\Gamma^2)}\, {(i\epsilon_0\,{\bf{E}}\times{\bf{E}}^\star)}\frac{\omega}{4 (k_B T)^2}&\propto& \omega^{-1} \,\,\,\,\, ;\,\,\,\,\,\beta\Gamma< \beta\omega \ll 1\\
    \nonumber
    = \mu_B \frac{\omega_p^2}{2\omega(\omega^2 + 16\Gamma^2)}\, {(i\epsilon_0\,{\bf{E}}\times{\bf{E}}^\star)}\,\,\,\,\,\,\,\,\,\,\,\,\,\,\,\,\,\,  &\propto& \omega^{-3} \,\,\,\,\,;\,\,\,\,\,\beta\Gamma \ll 1 \lesssim \beta\omega\,\,\,.
\end{eqnarray}
For strong disorder 
set by $\beta\Gamma \gtrsim 1$,
\begin{eqnarray}
\label{Unitary}
    {\bf{M}}_{ind} = \mu_B \frac{\omega_p^2}{2(\omega^2 + 16\Gamma^2)}\, {(i\epsilon_0\,{\bf{E}}\times{\bf{E}}^\star)}\frac{\omega}{4 (k_B T)^2} &\propto& \omega \,\,\,\,\,\,\,\,\,\,;\,\,\,\,\,\beta\omega \ll 1 \lesssim \beta\Gamma \\
    \nonumber
    = \mu_B \frac{\omega_p^2}{2\omega(\omega^2 + 16\Gamma^2)}\, {(i\epsilon_0\,{\bf{E}}\times{\bf{E}}^\star)} \,\,\,\,\,\,\,\,\,\,\,\,\,\,\,\,\,\,&\propto& \omega^{-1} \,\,\,\,\,;\,\,\,\,\,1 \ll \beta\omega \lesssim \beta\Gamma \\
    \nonumber
    = \mu_B \frac{\omega_p^2}{2\omega(\omega^2 + 16\Gamma^2)}\, {(i\epsilon_0\,{\bf{E}}\times{\bf{E}}^\star)}\,\,\,\,\,\,\,\,\,\,\,\,\,\,\,\,\,\, &\propto& \omega^{-3} \,\,\,\,\,;\,\,\,\,\,1 \lesssim  \beta\Gamma \ll \beta\omega\,\,\,.
\end{eqnarray}
The induced magnetization peaks at $\omega\sim\Gamma$ for weak disorder ($\Gamma \lesssim k_B T$) and at $\omega\sim k_B T$ for strong disorder($\Gamma \gtrsim k_B T$). In the low-frequency limit, screening effects become important. We have not included these effects in our calculation as we focus on the optical response. We note that the expressions above are for the local magnetization density ${\bf{M}}_{ind}$. A direct measurement of the total magnetization induced (for example, over the spot size of the light beam) would be an integral of the magnetization density over the pulse volume. This is important in the context of field inhomogeneity set by light of arbitrary pulse shapes.
The effective induced field is given by ${\bf{B}}_{eff} = \mu_0\,{\bf{M}}_{ind}$.
For metallic $Nb$, the induced fields via the inverse Faraday effect are plotted in Fig \ref{fig:B-eff}. Table I lists the effective induced fields at room temperature for some materials.

\begin{table}[h]
    \centering
    \begin{tabular}{c | c| c| c |c |c}
       Material  &  $\tau (\times10^{-12} sec)$ &$\omega_p$ (eV)  & $B_{eff}$(T) & $\omega(THz)$ & T(K)\\
       \hline
    Nb &  0.01 & 20 & 0.16 & 1 & 10\\
    Pb &  0.02 & 1.8 & 0.04 & 1 & 10 \\
    Pt & 0.02 & 7.5 & 0.005 & 1 & 10\\
    Au NP  & 0.0273 & 5.8 & 0.04 & $10^3$ & 300\\
    \label{tab:M-ind}
    \end{tabular}
    \caption{Effective induced magnetic fields at $T\sim 10 K$ for various materials in units of Tesla. $\tau$ is the relaxation time for electrons. $E=100 kV/cm$ for $\omega=1 THz$ was used ~\cite{BonettiNatPhys:2019}.  The values of the entries for Au nanoparticles are taken from ~\cite{NatPhotonics:2020} with laser peak intensity $9.1\times 10^{13} W/m^3$.}
\end{table}

The metallic skin depth of metals is in the range of $\lesssim 100nm$ and the induced magnetic moments in metals can be measured in finite size geometries. Indeed, this effect has been calculated using {\it{ab initio}} methods ~\cite{Oppeneer:2018} and observed in Au nanoparticles ~\cite{NatPhotonics:2020}.
Our calculation yields a result(Table I) consistent with  the measured values of effective fields induced via the IFE in Au nanoparticles reported in \cite{NatPhotonics:2020}, and consistent with numerical {\it{ab initio}} predictions reported in \cite{Oppeneer:2018}.

The induced fields predicted by equation(\ref{M-ind}) and listed in Table I are effected by photoinduced orbital magnetic moments of electrons in a metal, in the absence of SO coupling. We compare this to the induced fields effected by photoinduced spin moments of electrons, in the presence of SO coupling. For Pt, \cite{MetalsTHzT0:2011} estimate spin-mediated $B_{eff} \sim 10^{-4} T$ for incident light characterised by $E = 10^7 V/m$, $\omega = 1THz$ at zero temperature. The effective induced orbital fields, listed for Pt in Table I, are much higher, by an order of magnitude, even at temperatures of $10 K$ (only increasing at lower temperatures as shown in Fig.\ref{fig:B-eff}), using the same light parameters as \cite{MetalsTHzT0:2011}. However, IFE-induced spin magnetization can be much larger than their orbital counterparts in ferromagnetic materials. \cite{SOsplitZeemantransfer:2016} predict an effective induced field of a few Tesla in Rashba ferromagnetic Co/Pt bilayers. This is two orders of magnitude larger than the orbitally induced fields via IFE in metals. For the ferromagnetic bilayer system, the frequency dependence of the spin-mediated $B_{eff}$  predicted in the Born limit at zero temperature \cite{SOsplitZeemantransfer:2016} is similar ($\propto 1/\omega$) to that in our equation(\ref{Born}) for weak disorder, though the exact functional form is sensitive to both temperature and disorder as such for the orbital case. The spin-mediated IFE is shown to yield larger effective induced fields in magnetic systems \cite{SOsplitZeemantransfer:2016}. The effects of intrinsic system magnetism on the  orbital counterpart are beyond the scope of this article and not discussed here.

\section{Summary}
We consider a Keldysh Green's function  approach to formulate the  pure optical generation of magnetic fields in metals via the IFE.  We perform a fully microscopic calculation that apllies to the entire temperature range  and  captures the effects of disorder scattering  with no phenomenological fitting parameters.  We give the estimated effective  magnetic fields induced for a range of materials in Table I. Our theory would be  formally applicable to the superconducting state where, with proper modifications of the equations of motion, we can explore the induction of  vortex states by locally induced magnetic fields via the IFE. The proposed effect  highlights the utility of so-called "rectification" for static manipulation of the phase of  quantum coherent states.

\section{Acknowledgements}
 We are grateful to G. Aeppli, S. Bonetti,  I. Khaymovich, P. Oppeneer, B. Spivak, O. Tjernberg, J. Wiesenrieder and J. A. Sauls for useful discussions.  We acknowledge support from European Research Council under the European Union Seventh Framework ERS-2018-SYG 810451 HERO,  the Knut and Alice Wallenberg Foundation KAW 2019.0068 and the University of Connecticut.

\appendix
\section{Quasiclassical Theory : Definitions}

The central object of the theory is the quasiclassical Green's function, $\check{g}$ which is the propagator for quasiparticles of effective mass $m^{\star}$ with energy $\varepsilon$ and Fermi momentum ${\bf{p}}_F = m^{\star} {\bf{v}}_F$, given by the solutions to the equation,
\begin{equation}
\label{Kitaeqn}
[\varepsilon\check{\tau}_3 - \check{\sigma},\check{g}]_{\circ} + i{\bf{v}}_F\cdot\partial_{\vec{R}}\check{g} 
= \check{0} \,\,\,,
\end{equation}
where 
\begin{equation}
\label{gNambu}
\check{g} = (\begin{array}{c c} \hat{g}^R & \hat{g}^K\\ \hat{0} & \hat{g}^A \end{array}) \,\,\,;\,\,\, \check{\tau}_3 = (\begin{array}{c c} \hat{\tau}_3 & \hat{0}\\ \hat{0} & \hat{\tau}_3 \end{array})\,\,\,;\,\,\,\check{0}= (\begin{array}{c c} \hat{0} & \hat{0}\\ \hat{0} & \hat{0} \end{array})\,\,\,.
\end{equation}
Here, $\hat{g}^{R,A,K}$ are the Retarded, Advanced and Keldysh propagators and the $\hat{\,}$ refers to matrices in Nambu space. $\hat{\tau}_{j}$ $(j = 1,2,3)$ are the Pauli matrices in Nambu space. $\check{\sigma}$ is the self-energy with corresponding Retarded, Advanced and Keldysh components. 
The circle product is defined as :
\begin{equation}
\label{circdefnew}
    \hat{a}\circ\hat{b}({\bf{p}},{\bf{R}},\varepsilon,t) = \hat{a}({\bf{p}},{\bf{R}},\varepsilon - \frac{1}{2i} \frac{\partial}{\partial t_2}, t_1) \, \hat{b}({\bf{p}},{\bf{R}}, \varepsilon + \frac{1}{2i} \frac{\partial}{\partial t_1}, t_2)  \mid_{t_1 = t_2 = t}\,\,\,,
\end{equation}
where the $\partial/\partial t_1$ should be understood as acting on the left ~\cite{RainerSerene:1983}.
 This formulation is applicable for arbitrary external frequencies, $\omega$ much smaller than the Fermi energy, $\omega\ll\varepsilon_F$. Putting equation(\ref{gNambu}) in equation(\ref{Kitaeqn}), we get separate equations for the $R,A,K$ propagators,
\begin{eqnarray}
\label{RAKeqns}
[\varepsilon\hat{\tau}_3 - \hat{\sigma}^{R,A},\hat{g}^{R,A}]_{\circ} + i{\bf{v}}_F\cdot\nabla\hat{g}^{R,A} 
&=& \hat{0}\,\,\,;\\
\nonumber
(\varepsilon\hat{\tau}_3 - \hat{\sigma}^R)\circ\hat{g}^K - \hat{\sigma}^K\circ\hat{g}^A + \hat{g}^R\circ\hat{\sigma}^K - \hat{g}^K\circ(\varepsilon\hat{\tau}_3 - \hat{\sigma}^A) +  i{\bf{v}}_F\cdot\nabla\hat{g}^K 
&=& \hat{0} \,\,\,.
\end{eqnarray}

Consider electromagnetic radiation of frequency $\omega$ shone on a metal. For high-frequency light, the primary response of electrons(quasiparticles) is to the oscillating electric field of the electromagnetic wave. For the purpose of this calculation, we neglect the response of electrons to the oscillating magnetic field of light. Consider the electric field to be given by ${\bf{E}} = {\bf{E}}_0 e^{i{\bf{K}}.{\bf{R}} - i\omega t}$. The response of quasiparticles to this field can be calculated using equations(\ref{RAKeqns}). We expand $\hat{g}$  in the external field ${\bf{E}}$,
\begin{equation}
\label{gexp}
\hat{g}^X = \hat{g}_0^X + \hat{g}_1^X + \hat{g}_2^X\,\,\,\,\,,(X = R,A,K)
\end{equation}
with $\hat{g}_i^X \propto \mathcal{O}({\bf{E}}^i)$ being the $i$-th order correction to $\hat{g}^X$ and evaluate the current response. The current density operator at a space-time point $({\bf{R}},t)$ is given by 
\begin{equation}
\label{j-defn}
{\bf{j}}({\bf{R}},t) = e \,n({\bf{R}},t)\, {\bf{v}}({\bf{R}},t) \,\,\,,
\end{equation}
where $n({\bf{R}},t)$ and ${\bf{v}}({\bf{R}},t)$ are the local density and velocity of charge carriers operators, respectively. (${\bf{R}}$ and $t$ are the centre of mass space and time coordinates).
\begin{eqnarray}
\label{nv}
    n({\bf{R}},t) &=&  \int \frac{d^3 p}{(2\pi)^3} n({\bf{p}},t) e^{i{\bf{p}}\cdot{\bf{R}} }\\
    \nonumber 
    {\bf{v}}({\bf{R}},t) &=&    \int \frac{d^3 p'}{(2\pi)^3} {\bf{v}}({\bf{p}}\,',t) e^{i{\bf{p}}\,'\cdot{\bf{R}}}
\end{eqnarray}
Using equations(\ref{nv}) in equation(\ref{j-defn}),  we get ${\bf{j}}$,
\begin{equation}
\label{j2n1v1}
    {\bf{j}}({\bf{R}},t) = e  \langle \,\bar{n}({\bf{p}})\,\, \bar{v}({\bf{p}}) \,\rangle_{FS}\,\,\,,
\end{equation}
where $\langle ...\rangle_{FS} \equiv \int \frac{d^3 p}{(2\pi)^3}$ refers to an average over the Fermi surface and 
\begin{eqnarray}
\label{n1v1}
    \bar{n}({\bf{p}}) &\equiv& N_f \int d\varepsilon\, \mathscr{T} \hat{g}^K({\bf{p}},\varepsilon, t) \\
    \nonumber 
    \bar{v}({\bf{p}})  &\equiv&   \frac{N_f}{n_0}{\bf{v}}_F \int {d\varepsilon'} \mathscr{T}\hat{\tau}_3 \hat{g}^K({\bf{p}},\varepsilon', t) \,\,\,.
\end{eqnarray}
Here, $n_0$ is the equilibrium density of charge carriers, $N_f$ is the density of states at the Fermi level and $\mathscr{T}$ refers to the trace over the Nambu and spin degrees of freedom.
 
At equilibrium, $\hat{g}_0$ is given by the solution to the Eilenberger equation in the absence of external fields,
\begin{equation}
\label{g0}
    \hat{g}_0^{R,A} = \pm i \, \hat{\tau}_3\,\,\,,\,\,\,\hat{g}_0^K = - 2 i\, tanh(\beta\varepsilon) \,\hat{\tau}_3\,\,\,,
\end{equation}
where $\beta\equiv (2 k_B T)^{-1}$ and $\varepsilon^{R,A} = \varepsilon \pm i0+$. 
We use the notation for the Nambu elements, $\hat{g}^X \equiv \Big( \begin{array}{c c} g^X & f^X\\ \underline{f}^X & \underline{g}^K\end{array}\Big)$. 
$g^X$ and $\underline{g}^X$ (and analogously $f^X$ and $\underline{f}^X$) are related by symmetries, $[\hat{g}^R]^{\dagger} = -\hat{\tau}_3\hat{g}^A\hat{\tau}_3$ and $[\hat{g}^K]^{\dagger} = \hat{\tau}_3\hat{g}^K\hat{\tau}_3$.

\section{Calculation of IFE}

We take into account the effect of impurities that scatter quasiparticles at a rate $\Gamma = \frac{\hbar}{\tau}$ where $\tau = \frac{l_{mfp}}{v_F}$ is the relaxation time and $l_{mfp}$ is the mean free path for scattering between point impurities. The dynamical response of electrons to light of a given frequency $\omega$ crosses over from the ballistic ($\Gamma\ll\omega$) to the diffusive ($\Gamma\gg\omega$) regime as the rate of impurity scattering increases. In the hydrodynamic limit, diffusive motion of electrons renders the propagators and transport equation  isotropic over the Fermi surface. The response of electrons in this diffusive limit is given by solutions to the Usadel equation for Fermi-surface averaged propagators. In the "clean" or ballistic limit, the propagators retain their momentum-space structure and are given by solutions to the Eilenberger equation(\ref{RAKeqns}). We consider the dynamics of electrons in this ballistic regime. The impurity self-energy is included as $\sigma^{R} = i\Gamma$. 

We are interested in calculating the time-averaged corrections to the current density ${\bf{j}}$ that are of second order in the field, ${\bf{j}}_2 \propto \mathcal{O}({\bf{E}}^2)$,
\begin{equation}
\label{j2}
{\bf{j}}_2 = e \langle n_2 {\bf{v}}_0 \rangle_{FS,(2\pi/\omega)} + e  \langle n_0 {\bf{v}}_2 \rangle_{FS,(2\pi/\omega)} + e  \langle n_1 {\bf{v}}_1 \rangle_{FS,(2\pi/\omega)} \,\,\,\,\,.
\end{equation}
$n_2({\bf{p}})$ is given by equation(\ref{n1v1}) using the second-order corrections $\hat{g}_2^K$. $\hat{g}_2^K$ is obtained from equation(\ref{RAKeqns}) with the driving term $\propto ({\bf{v}}_F\cdot{\bf{E}})\,\hat{g}_1^K$. As $\hat{g}_1^K \propto ({\bf{v}}_F\cdot{\bf{E}})$ (shown in equation(\ref{gKsolns})), this gives $\langle n_2 {\bf{v}}_0 \rangle_{FS} \propto \langle {\bf{v}}_F({\bf{v}}_F\cdot{\bf{E}})({\bf{v}}_F\cdot{\bf{E}}) \rangle_{FS} = 0$. Similarly, the second term in equation(\ref{j2}) also gives a zero contribution. This leaves only the third term non-zero in ${\bf{j}}_2$ above. $n_1$ and ${\bf{v}}_1$ are given by $\hat{g}_1^K$, as in equation(\ref{n1v1}). 

$\hat{g}_1^X$ are given by solutions to the equation,
\begin{eqnarray}
\label{RAKeqns2}
 (\varepsilon^{R,A}\,\hat{\tau}_3 - \hat{\sigma}^{R,A} - \hat{\sigma}_{ext}^{R,A})\circ\hat{g}^{R,A} - \hat{g}^{R,A}\circ(\varepsilon^{R,A}\,\hat{\tau}_3 - \hat{\sigma}^{R,A} - \hat{\sigma}_{ext}^{R,A}) + i{\bf{v}}_F\cdot\nabla\hat{g}^{R,A} = 0 &\,&\\
 \nonumber
 (\varepsilon^{R}\,\hat{\tau}_3 - \hat{\sigma}^{R} - \hat{\sigma}_{ext}^R)\circ\hat{g}^K - \hat{g}^K\circ(\varepsilon^{A}\,\hat{\tau}_3 - \hat{\sigma}^{A} - \hat{\sigma}_{ext}^A) + \hat{g}^R\circ\hat{\sigma}^K -\hat{\sigma}^K\circ\hat{g}^A + i{\bf{v}}_F\cdot\nabla\hat{g}^K  &=& 0\,,
\end{eqnarray}
where $\hat{\sigma}_{ext}$ is the applied external field. 
Using the notation in equation(\ref{gexp}), the definition(\ref{circdefnew}), and the self-energies given by $\hat{\sigma}^{R,A} =\pm i\Gamma\hat{\tau}_3$, we evaluate the convolution products. With 
\begin{equation}
\label{sigma0}
    \hat{\sigma}_0^K = (\hat{\sigma}_0^R - \hat{\sigma}_0^A)\, tanh(\beta\varepsilon) = 2i\Gamma \hat{\tau}_3\, tanh(\beta\varepsilon)\,\,\,,
\end{equation} 
and
\begin{equation}
\label{sigma1}
    \hat{\sigma}_1^K = \Gamma \hat{t}^R \hat{g}^K \hat{t}^A = \Gamma\hat{g}^K\,\,\,,
\end{equation}
the driving terms given by $\hat{\sigma}_{ext}^{R,A}\circ\hat{g}^{R,A}- \hat{g}^{R,A}\circ\hat{\sigma}_{ext}^{R,A}$ for the Retarded/Advanced components  and $\hat{\sigma}_{ext}^{R}\circ\hat{g}^K - \hat{g}^K\circ\hat{\sigma}_{ext}^A $ for the $K$ component, respectively. The applied external field $({\bf{A}},\Phi)$ gives the associated self-energy, $\hat{\sigma}_{ext} = e{\bf{v}}_F\cdot{\bf{A}} - e\Phi$. The vector potential ${\bf{A}}$ is related to the oscillating electric field of light, ${\bf{E}} = \frac{({\bf{E}}_0\,e^{-i\omega t} + c.c.)}{2} = -\frac{\partial {A}}{\partial t}$.
We neglect screening effects and set $\Phi = 0$. Then to $\mathcal{O}({\bf{E}})$,
,
\begin{eqnarray}
\label{MatrixEqnsgRAK}
    (\varepsilon - i\Gamma - \frac{1}{2i}\frac{\partial}{\partial t})\,\hat{\tau}_3\hat{g}_1^K &-& (\varepsilon + i\Gamma + \frac{1}{2i}\frac{\partial}{\partial t})\,\hat{g}_1^K\hat{\tau}_3\\
    \nonumber
    -2i\Gamma\,tanh\Big(\beta(\varepsilon - \frac{1}{2i}\frac{\partial}{\partial t})\Big)\,\hat{\tau}_3\hat{g}_1^A &+& 2i\Gamma\,tanh\Big(\beta(\varepsilon + \frac{1}{2i}\frac{\partial}{\partial t})\Big)\,\hat{g}_1^R\hat{\tau}_3\\
    \nonumber
- \Gamma\,\hat{g}_1^K\,\hat{g}_0^A(\varepsilon + \frac{1}{2i}\frac{\partial}{\partial t}) &+& \hat{g}_0^R(\varepsilon - \frac{1}{2i}\frac{\partial}{\partial t})\Gamma\,\hat{g}_1^K\\
    \nonumber
    + i{\bf{v}}_F\cdot\nabla \hat{g}_1^K + 2 i\,(tanh\Big(\beta(\varepsilon + \frac{\omega}{2})\Big) &-& tanh\Big(\beta(\varepsilon - \frac{\omega}{2})\Big))\,\frac{e}{-i\omega}{\bf{v}}_F\cdot{\bf{E}} = 0\,\,\,,\\
    \nonumber
    (\varepsilon - i\Gamma - \frac{1}{2i}\frac{\partial}{\partial t})\,\hat{\tau}_3\hat{g}_1^R - (\varepsilon - i\Gamma &+& \frac{1}{2i}\frac{\partial}{\partial t})\,\hat{g}_1^R\hat{\tau}_3 + i{\bf{v}}_F\cdot\nabla \hat{g}_1^R   = 0\,\,\,.
\end{eqnarray}
Examining these equations,  we seek solutions of the form
\begin{equation}
\label{g-ansatz}
    \hat{g}_1^X = \hat{g}_+^X\,e^{i\omega t} + \hat{g}_-^X\,e^{-i\omega t}\,\,\,.
\end{equation}
$\hat{g}_\pm^X = (\begin{array}{c c} g_\pm^X & 0\\ 0 & \underline{g}^X_\pm \end{array})$, using the notation defined in Appendix A, with the symmetry $\underline{g}^R(\hat{p},\varepsilon) = -g^R(-\hat{p},-\varepsilon)^{\star}$ and $\underline{g}^K(\hat{p},\varepsilon) = g^K(-\hat{p},-\varepsilon)^{\star}$. 
(Here, ${\bf{p}} = p_F\hat{p}$). 
Using the ansatz, equation(\ref{g-ansatz}) in the Retarded equation(\ref{MatrixEqnsgRAK}), 
we get equations for $g_\pm^R$ ($\underline{g}_\pm$ can be obtained using symmetries),
\begin{equation}
    (\varepsilon - i\Gamma \pm \frac{\omega}{2})\,g_\mp^R - (\varepsilon - i\Gamma \mp \frac{\omega}{2})\,g_\mp^R + i{\bf{v}}_F\cdot\nabla g_\mp^R  
    = 0\,\,\,.
    \end{equation}
Now ${\bf{v}}_F\cdot{\bf{K}} \propto \frac{v_F}{c}\omega \ll \omega$. So we use ${\bf{K}}=0$ solutions. From the equation above,
\begin{equation}
\label{gR}
    g_\pm^{R} = \underline{g}_\mp^R = 0\,\,\,.
\end{equation}
Similarly, $\hat{g}_\pm^A = 0$.
Finally, for the Keldysh components,
\begin{eqnarray}
\label{gKeqns}
\nonumber
(\varepsilon - i\Gamma +\frac{\omega}{2})\,g_-^K &-& (\varepsilon + i\Gamma - \frac{\omega}{2})\,g_-^K - 2i\Gamma\,tanh\Big(\beta(\varepsilon +\frac{\omega}{2})\Big)\,g_-^A\\
+ 2i\Gamma tanh\Big(\beta(\varepsilon - \frac{\omega}{2})\Big)\,g_-^R
&-& i\Gamma\,g_-^K\, - i\Gamma\,g_-^K
+ i{\bf{v}}_F\cdot\nabla g_-^K \\
\nonumber
+ 2 i\,(tanh\Big(\beta(\varepsilon &+& \frac{\omega}{2})\Big) - tanh\Big(\beta(\varepsilon - \frac{\omega}{2})\Big))\,{\frac{e}{-2i\omega}}{\bf{v}}_F\cdot{\bf{E}}_0 = 0 \,,\\
\nonumber
(\varepsilon - i\Gamma - \frac{\omega}{2})\,g_+^K &-& (\varepsilon + i\Gamma + \frac{\omega}{2})\,g_+^K - 2i\Gamma\,tanh\Big(\beta(\varepsilon -\frac{\omega}{2})\Big)\,g_+^A\\
+ 2i\Gamma tanh\Big(\beta(\varepsilon + \frac{\omega}{2})\Big)\,g_+^R
&-& i\Gamma\,g_+^K\, - i\Gamma\,g_+^K
+ i{\bf{v}}_F\cdot\nabla g_+^K \\
\nonumber
+ 2 i\,(tanh\Big(\beta(\varepsilon &-& \frac{\omega}{2})\Big) - tanh\Big(\beta(\varepsilon + \frac{\omega}{2})\Big))\,{\frac{e}{2i\omega}}{\bf{v}}_F\cdot{\bf{E}}_0^\star = 0 \,,
\end{eqnarray}
and using the R,A solutions in  equations(\ref{gR}), the ${\bf{K}}=0$ solutions are
\begin{eqnarray}
\label{gKsolns}
g_-^K &=& \frac{ e}{\omega(\omega - 4i\Gamma)}({\bf{v}}_F\cdot{\bf{E}}_0) (tanh\Big(\beta(\varepsilon + \frac{\omega}{2})\Big) - tanh\Big(\beta(\varepsilon - \frac{\omega}{2})\Big))\\
\nonumber
\underline{g}_+^K &=& \frac{ e}{\omega(\omega + 4i\Gamma)}({\bf{v}}_F\cdot{\bf{E}}_0^\star) (tanh\Big(\beta(\varepsilon + \frac{\omega}{2})\Big) - tanh\Big(\beta(\varepsilon - \frac{\omega}{2})\Big))\,\,\,,\\
\nonumber
g_+^K &=& - \frac{ e}{\omega(\omega + 4i\Gamma)}({\bf{v}}_F\cdot{\bf{E}}_0^\star) (tanh\Big(\beta(\varepsilon + \frac{\omega}{2})\Big) - tanh\Big(\beta(\varepsilon - \frac{\omega}{2})\Big))\\
\nonumber
\underline{g}_+^K &=& - \frac{ e}{\omega(\omega - 4i\Gamma)}({\bf{v}}_F\cdot{\bf{E}}_0) (tanh\Big(\beta(\varepsilon + \frac{\omega}{2})\Big) - tanh\Big(\beta(\varepsilon - \frac{\omega}{2})\Big))\,\,\,.
\end{eqnarray}
Using the ansatz in equation(\ref{g-ansatz}) to find $n_1$ and ${\bf{v}}_1$, given by equation(\ref{n1v1}),
\begin{eqnarray}
\label{n1vi-gKs}
    n_1 &=& N_f \int{d\varepsilon}\, (g^K + \underline{g}^K)\\
    \nonumber
    &=& N_f \int {d\varepsilon}\, ( g_-^K\,e^{-i\omega t} + \underline{g}_+^K\, e^{i\omega t} + g_+^K\,e^{i\omega t} + \underline{g}_-^K\, e^{-i\omega t} )\\
    \nonumber
    {\bf{v}}_1 &=& \frac{N_f }{n_0} {\bf{v}}_F\int{d\varepsilon}\, (g_-^K\,e^{-i\omega t} + g_+^K\,e^{i\omega t} - \underline{g}_+^K\, e^{i\omega t} - \underline{g}_-^K\, e^{-i\omega t} )\,\,\,.
\end{eqnarray}
In the limit ${\bf{K}} = 0$, and in the absence of all self-energy and applied external field terms, equations(\ref{gKeqns}) give
\begin{eqnarray}
\label{ggrad}
    \varepsilon\circ {g}_\pm^K - {g}_\pm^K\circ\varepsilon + i{\bf{v}}_F\cdot\nabla g_\pm^K &=& 0\\
    \nonumber
    g_\pm^K &=& - \frac{i{\bf{v}}_F\cdot\nabla g_\pm^K}{\omega}\,\,\,.
\end{eqnarray}
This is an expansion of $g_\pm^K$ valid to $\mathcal{O}({\bf{E}})$ and to leading order in the gradient $\mathcal{O}(\nabla)$. $g_\pm^K$ is of order $\mathcal{O}({\bf{E}})$ and we have used equation(\ref{ggrad}) for free electrons (zeroth order in self-energies) to obtain the $\nabla$ term, which is therefore of order $\mathcal{O}({\bf{E}}^0)$, thus giving the expansion(\ref{ggrad}) of $\mathcal{O}({\bf{E}})$.
Now, define
\begin{equation}
\label{BigG}
\mathcal{G}_\pm^X \equiv \int d\varepsilon \,g_\pm^X\,\,\,;\,\,\,\underline{\mathcal{G}}_\pm^X \equiv \int d\varepsilon \,\underline{g}_\pm^X\,\,\,,
\end{equation}
and use equations(\ref{n1vi-gKs}-\ref{ggrad}) in equation(\ref{j2n1v1}) to find (to leading order in $\mathcal{O}(\nabla)$),
\begin{eqnarray}
{\bf{j}}_2 &=& e \frac{N_f^2}{ n_0}\,\langle \,{\bf{v}}_F (\mathcal{G}_-^K\,e^{-i\omega t} + \mathcal{G}_+^K\,e^{i\omega t} + \underline{\mathcal{G}}_+^K\, e^{i\omega t} + \underline{\mathcal{G}}_-^K\, e^{-i\omega t} ) \,\,\,\\
\nonumber
&\,&\cdot\{- \frac{i{\bf{v}}_F\cdot\nabla}{\omega}( \mathcal{G}_-^K\,e^{-i\omega t} + \mathcal{G}_+^K\,e^{i\omega t} - \underline{\mathcal{G}}_+^K\, e^{i\omega t} - \underline{\mathcal{G}}_-^K\, e^{-i\omega t})\}\,\rangle_{FS}\,\,\,.
\end{eqnarray}
Time-averaging over a period of oscillation, we get static terms,
\begin{equation}
\label{j2tavg}
\langle{\bf{j}}_2\rangle_{2\pi/\omega} =  e \frac{N_f^2}{ \omega \,n_0}\,\langle \,{\bf{v}}_F ( \mathcal{G}_-^K (i{\bf{v}}_F\cdot\nabla) \underline{\mathcal{G}}_+^K - \underline{\mathcal{G}}_+^K(i{\bf{v}}_F\cdot\nabla)\mathcal{G}_-^K + \mathcal{G}_+^K (i{\bf{v}}_F\cdot\nabla) \underline{\mathcal{G}}_-^K - \underline{\mathcal{G}}_-^K(i{\bf{v}}_F\cdot\nabla)\mathcal{G}_+^K )\rangle_{FS}\,\,\,.
\end{equation}
Using the definitions(\ref{BigG}) and the solutions (\ref{gKsolns}) to evaluate $\mathcal{G}^K$,
\begin{eqnarray}
    \mathcal{G}_-^K &=& \frac{ e}{2(\omega - 4i\Gamma)}({\bf{v}}_F\cdot{\bf{E}}_0)  \zeta(\beta\omega)\,\,\,,\\
\nonumber
\underline{\mathcal{G}}_+^K &=& \frac{ e}{2(\omega + 4i\Gamma)}({\bf{v}}_F\cdot{\bf{E}}_0^\star) \zeta(\beta\omega)\,\,\,,\\
\nonumber
\mathcal{G}_+^K &=& -\frac{ e}{2(\omega + 4i\Gamma)}({\bf{v}}_F\cdot{\bf{E}}_0^\star)  \zeta(\beta\omega)\,\,\,,\\
\nonumber
\underline{\mathcal{G}}_-^K &=& - \frac{ e}{2(\omega - 4i\Gamma)}({\bf{v}}_F\cdot{\bf{E}}_0) \zeta(\beta\omega)\,\,\,,
\end{eqnarray}
where 
\begin{equation}
\zeta(\beta\omega) =   \frac{tanh(\beta\omega)}{\sqrt{2 - tanh^2(\beta\omega)}}\,.
\end{equation}
We now use these in the expression(\ref{j2tavg}) to get the static induced current density. 
\begin{equation}
\langle{\bf{j}}_2\rangle_{2\pi/\omega} =    \frac{i\,e^3 N_f^2\,\zeta^2(\beta\omega)}{ \omega (\omega^2 + 16\Gamma^2)n_0}\langle\,{\bf{v}}_F({\bf{v}}_F\cdot{\bf{E}}_0)({\bf{v}}_F\cdot\nabla)({\bf{v}}_F\cdot{\bf{E}}_0^\star) - {\bf{v}}_F({\bf{v}}_F\cdot{\bf{E}}_0^\star)({\bf{v}}_F\cdot\nabla)({\bf{v}}_F\cdot{\bf{E}}_0)\rangle_{FS}\,.
\end{equation}
 Evaluating the Fermi surface averages, 
\begin{equation}
\langle{\bf{j}}_2\rangle_{2\pi/\omega} = \frac{ e}{4m} \frac{\omega_p^2}{(\omega^2 + 16\Gamma^2)} \frac{\nabla\times(i\epsilon_0\, {\bf{E}}\times{\bf{E}}^\star)}{\omega}\zeta^2(\beta\omega) + {\bf{j}}_{pm}\,\,\,,
\end{equation}
since $N_F\varepsilon_F = n_0$ and the plasma frequency $\omega_p = n_0 e^2/m \epsilon_0$. The second term refers to the current induced by ponderomotive forces on the electrons. The first term above may be expressed in terms of a magnetization density, ${\bf{M}}_{ind}$ induced via the IFE viz.,
\begin{eqnarray}
\nonumber
    {\bf{j}}_{IFE} &=& \nabla \times {\bf{M}}_{ind}\\
    {\bf{M}}_{ind} &=& \mu_B \frac{\omega_p^2}{2\omega(\omega^2 + 16\Gamma^2)}\, {(i\epsilon_0\,{\bf{E}}\times{\bf{E}}^\star)}\zeta^2(\beta\omega)\,\,\,.
    \label{Hertel}
\end{eqnarray}

\bibliography{IFE_Metal/Paper_IFEMetal}

\begin{thebibliography}{33}%
\makeatletter
\providecommand \@ifxundefined [1]{%
 \@ifx{#1\undefined}
}%
\providecommand \@ifnum [1]{%
 \ifnum #1\expandafter \@firstoftwo
 \else \expandafter \@secondoftwo
 \fi
}%
\providecommand \@ifx [1]{%
 \ifx #1\expandafter \@firstoftwo
 \else \expandafter \@secondoftwo
 \fi
}%
\providecommand \natexlab [1]{#1}%
\providecommand \enquote  [1]{``#1''}%
\providecommand \bibnamefont  [1]{#1}%
\providecommand \bibfnamefont [1]{#1}%
\providecommand \citenamefont [1]{#1}%
\providecommand \href@noop [0]{\@secondoftwo}%
\providecommand \href [0]{\begingroup \@sanitize@url \@href}%
\providecommand \@href[1]{\@@startlink{#1}\@@href}%
\providecommand \@@href[1]{\endgroup#1\@@endlink}%
\providecommand \@sanitize@url [0]{\catcode `\\12\catcode `\$12\catcode
  `\&12\catcode `\#12\catcode `\^12\catcode `\_12\catcode `\%12\relax}%
\providecommand \@@startlink[1]{}%
\providecommand \@@endlink[0]{}%
\providecommand \url  [0]{\begingroup\@sanitize@url \@url }%
\providecommand \@url [1]{\endgroup\@href {#1}{\urlprefix }}%
\providecommand \urlprefix  [0]{URL }%
\providecommand \Eprint [0]{\href }%
\providecommand \doibase [0]{https://doi.org/}%
\providecommand \selectlanguage [0]{\@gobble}%
\providecommand \bibinfo  [0]{\@secondoftwo}%
\providecommand \bibfield  [0]{\@secondoftwo}%
\providecommand \translation [1]{[#1]}%
\providecommand \BibitemOpen [0]{}%
\providecommand \bibitemStop [0]{}%
\providecommand \bibitemNoStop [0]{.\EOS\space}%
\providecommand \EOS [0]{\spacefactor3000\relax}%
\providecommand \BibitemShut  [1]{\csname bibitem#1\endcsname}%
\let\auto@bib@innerbib\@empty
\bibitem [{\citenamefont {Cheng}\ \emph {et~al.}(2020)\citenamefont {Cheng},
  \citenamefont {Son},\ and\ \citenamefont {Sheldon}}]{NatPhotonics:2020}%
  \BibitemOpen
  \bibfield  {author} {\bibinfo {author} {\bibfnamefont {O.~H.-C.}\
  \bibnamefont {Cheng}}, \bibinfo {author} {\bibfnamefont {D.~H.}\ \bibnamefont
  {Son}},\ and\ \bibinfo {author} {\bibfnamefont {M.}~\bibnamefont {Sheldon}},\
  }\bibfield  {title} {\bibinfo {title} {{Light-induced magnetism in plasmonic
  gold nanoparticles}},\ }\href
  {https://www.nature.com/articles/s41566-020-0603-3#Abs1} {\bibfield
  {journal} {\bibinfo  {journal} {Nature Photonics}\ }\textbf {\bibinfo
  {volume} {14}},\ \bibinfo {pages} {365} (\bibinfo {year} {2020})}\BibitemShut
  {NoStop}%
\bibitem [{\citenamefont {Hurst}\ \emph {et~al.}(2018)\citenamefont {Hurst},
  \citenamefont {Oppeneer}, \citenamefont {Manfredi},\ and\ \citenamefont
  {Hervieux}}]{Oppeneer:2018}%
  \BibitemOpen
  \bibfield  {author} {\bibinfo {author} {\bibfnamefont {J.}~\bibnamefont
  {Hurst}}, \bibinfo {author} {\bibfnamefont {P.~M.}\ \bibnamefont {Oppeneer}},
  \bibinfo {author} {\bibfnamefont {G.}~\bibnamefont {Manfredi}},\ and\
  \bibinfo {author} {\bibfnamefont {P.-A.}\ \bibnamefont {Hervieux}},\
  }\bibfield  {title} {\bibinfo {title} {{Magnetic moment generation in small
  gold nanoparticles via the plasmonic inverse Faraday effect}},\ }\href
  {https://journals.aps.org/prb/abstract/10.1103/PhysRevB.98.134439} {\bibfield
   {journal} {\bibinfo  {journal} {Phys. Rev. B}\ }\textbf {\bibinfo {volume}
  {98}},\ \bibinfo {pages} {134439} (\bibinfo {year} {2018})}\BibitemShut
  {NoStop}%
\bibitem [{\citenamefont {Shaw}\ \emph {et~al.}(2019)\citenamefont {Shaw},
  \citenamefont {Blanco~Alvarez}, \citenamefont {Brisbois}, \citenamefont
  {Burger}, \citenamefont {Pinheiro}, \citenamefont {Kramer}, \citenamefont
  {Motta}, \citenamefont {Fleury-Frenette}, \citenamefont {Ortiz},
  \citenamefont {Vanderheyden},\ and\ \citenamefont {Silhanek}}]{Metals:2019}%
  \BibitemOpen
  \bibfield  {author} {\bibinfo {author} {\bibfnamefont {G.}~\bibnamefont
  {Shaw}}, \bibinfo {author} {\bibfnamefont {S.}~\bibnamefont
  {Blanco~Alvarez}}, \bibinfo {author} {\bibfnamefont {J.}~\bibnamefont
  {Brisbois}}, \bibinfo {author} {\bibfnamefont {L.}~\bibnamefont {Burger}},
  \bibinfo {author} {\bibfnamefont {L.~B.}\ \bibnamefont {Pinheiro}}, \bibinfo
  {author} {\bibfnamefont {R.~B.}\ \bibnamefont {Kramer}}, \bibinfo {author}
  {\bibfnamefont {M.}~\bibnamefont {Motta}}, \bibinfo {author} {\bibfnamefont
  {K.}~\bibnamefont {Fleury-Frenette}}, \bibinfo {author} {\bibfnamefont
  {W.}~\bibnamefont {Ortiz}}, \bibinfo {author} {\bibfnamefont
  {B.}~\bibnamefont {Vanderheyden}},\ and\ \bibinfo {author} {\bibfnamefont
  {A.}~\bibnamefont {Silhanek}},\ }\bibfield  {title} {\bibinfo {title}
  {{Magnetic Recording of Superconducting States}},\ }\href
  {https://doi.org/10.3390/met9101022} {\bibfield  {journal} {\bibinfo
  {journal} {Metals}\ }\textbf {\bibinfo {volume} {9}},\ \bibinfo {pages}
  {1022} (\bibinfo {year} {2019})}\BibitemShut {NoStop}%
\bibitem [{\citenamefont {Wang}\ \emph {et~al.}(2022)\citenamefont {Wang},
  \citenamefont {Xiao}, \citenamefont {Park}, \citenamefont {Zhu},
  \citenamefont {WangM}, \citenamefont {Taniguchi}, \citenamefont {Watanabe},
  \citenamefont {Yan}, \citenamefont {Xiao}, \citenamefont {Gamelin},
  \citenamefont {Yao},\ and\ \citenamefont {Xu}}]{FerromagMoire:2022}%
  \BibitemOpen
  \bibfield  {author} {\bibinfo {author} {\bibfnamefont {X.}~\bibnamefont
  {Wang}}, \bibinfo {author} {\bibfnamefont {C.}~\bibnamefont {Xiao}}, \bibinfo
  {author} {\bibfnamefont {H.}~\bibnamefont {Park}}, \bibinfo {author}
  {\bibfnamefont {J.}~\bibnamefont {Zhu}}, \bibinfo {author} {\bibfnamefont
  {C.}~\bibnamefont {WangM}}, \bibinfo {author} {\bibfnamefont
  {T.}~\bibnamefont {Taniguchi}}, \bibinfo {author} {\bibfnamefont
  {K.}~\bibnamefont {Watanabe}}, \bibinfo {author} {\bibfnamefont
  {J.}~\bibnamefont {Yan}}, \bibinfo {author} {\bibfnamefont {D.}~\bibnamefont
  {Xiao}}, \bibinfo {author} {\bibfnamefont {D.~R.}\ \bibnamefont {Gamelin}},
  \bibinfo {author} {\bibfnamefont {W.}~\bibnamefont {Yao}},\ and\ \bibinfo
  {author} {\bibfnamefont {X.}~\bibnamefont {Xu}},\ }\bibfield  {title}
  {\bibinfo {title} {{Light-induced Ferromagnetism in Moirè Superlattices}},\
  }\href@noop {} {\bibfield  {journal} {\bibinfo  {journal} {Nature}\ }\textbf
  {\bibinfo {volume} {604}},\ \bibinfo {pages} {468} (\bibinfo {year}
  {2022})}\BibitemShut {NoStop}%
\bibitem [{\citenamefont {Fausti}\ \emph {et~al.}(2011)\citenamefont {Fausti},
  \citenamefont {Tobey}, \citenamefont {Dean}, \citenamefont {Kaiser},
  \citenamefont {Dienst}, \citenamefont {Hoffmann}, \citenamefont {Pyon},
  \citenamefont {Takayama}, \citenamefont {Takagi},\ and\ \citenamefont
  {Cavalleri}}]{SCCuprate:2011}%
  \BibitemOpen
  \bibfield  {author} {\bibinfo {author} {\bibfnamefont {D.}~\bibnamefont
  {Fausti}}, \bibinfo {author} {\bibfnamefont {R.}~\bibnamefont {Tobey}},
  \bibinfo {author} {\bibfnamefont {N.}~\bibnamefont {Dean}}, \bibinfo {author}
  {\bibfnamefont {S.}~\bibnamefont {Kaiser}}, \bibinfo {author} {\bibfnamefont
  {A.}~\bibnamefont {Dienst}}, \bibinfo {author} {\bibfnamefont {M.~C.}\
  \bibnamefont {Hoffmann}}, \bibinfo {author} {\bibfnamefont {S.}~\bibnamefont
  {Pyon}}, \bibinfo {author} {\bibfnamefont {T.}~\bibnamefont {Takayama}},
  \bibinfo {author} {\bibfnamefont {H.}~\bibnamefont {Takagi}},\ and\ \bibinfo
  {author} {\bibfnamefont {A.}~\bibnamefont {Cavalleri}},\ }\bibfield  {title}
  {\bibinfo {title} {Light-induced superconductivity in a stripe-ordered
  cuprate},\ }\href@noop {} {\bibfield  {journal} {\bibinfo  {journal}
  {Science}\ }\textbf {\bibinfo {volume} {331}},\ \bibinfo {pages} {189}
  (\bibinfo {year} {2011})}\BibitemShut {NoStop}%
\bibitem [{\citenamefont {Kogar}\ \emph {et~al.}(2020)\citenamefont {Kogar}
  \emph {et~al.}}]{CDWNatPhys:2020}%
  \BibitemOpen
  \bibfield  {author} {\bibinfo {author} {\bibfnamefont {A.}~\bibnamefont
  {Kogar}} \emph {et~al.},\ }\bibfield  {title} {\bibinfo {title}
  {Light-induced charge density wave in \begin{math}{La Te_3}\end{math}},\
  }\href@noop {} {\bibfield  {journal} {\bibinfo  {journal} {Nat. Phys.}\
  }\textbf {\bibinfo {volume} {16}},\ \bibinfo {pages} {159} (\bibinfo {year}
  {2020})}\BibitemShut {NoStop}%
\bibitem [{\citenamefont {Kimel}\ \emph {et~al.}(2022)\citenamefont {Kimel}
  \emph {et~al.}}]{RoadMap:2022}%
  \BibitemOpen
  \bibfield  {author} {\bibinfo {author} {\bibfnamefont {A.}~\bibnamefont
  {Kimel}} \emph {et~al.},\ }\bibfield  {title} {\bibinfo {title} {{2022
  Magneto-Optics RoadMap}},\ }\href@noop {} {\bibfield  {journal} {\bibinfo
  {journal} {J. Phys. D: Appl. Phys.}\ }\textbf {\bibinfo {volume} {55}},\
  \bibinfo {pages} {463003} (\bibinfo {year} {2022})}\BibitemShut {NoStop}%
\bibitem [{\citenamefont {Croitoru}\ \emph
  {et~al.}(2022{\natexlab{a}})\citenamefont {Croitoru}, \citenamefont
  {Mironov}, \citenamefont {Lounis},\ and\ \citenamefont
  {Buzdin}}]{AdvQuantTech:2022}%
  \BibitemOpen
  \bibfield  {author} {\bibinfo {author} {\bibfnamefont {M.~D.}\ \bibnamefont
  {Croitoru}}, \bibinfo {author} {\bibfnamefont {S.~V.}\ \bibnamefont
  {Mironov}}, \bibinfo {author} {\bibfnamefont {B.}~\bibnamefont {Lounis}},\
  and\ \bibinfo {author} {\bibfnamefont {A.~I.}\ \bibnamefont {Buzdin}},\
  }\bibfield  {title} {\bibinfo {title} {{Toward the Light-Operated
  Superconducting Devices: Circularly Polarized Radiation Manipulates the
  Current-Carrying States in Superconducting Rings}},\ }\href
  {https://doi.org/10.1002/qute.202200054} {\bibfield  {journal} {\bibinfo
  {journal} {Advanced Quantum Technologies}\ }\textbf {\bibinfo {volume}
  {82}},\ \bibinfo {pages} {2200054} (\bibinfo {year}
  {2022}{\natexlab{a}})}\BibitemShut {NoStop}%
\bibitem [{\citenamefont {Chen}(2017)}]{Sc.Reps:2017}%
  \BibitemOpen
  \bibfield  {author} {\bibinfo {author} {\bibfnamefont {X.-J.}\ \bibnamefont
  {Chen}},\ }\bibfield  {title} {\bibinfo {title} {{Fundamental mechanism for
  all-optical helicity-dependent switching of magnetization}},\ }\href
  {https://www.nature.com/articles/srep41294#Bib1} {\bibfield  {journal}
  {\bibinfo  {journal} {Sc. Reps.}\ }\textbf {\bibinfo {volume} {7}},\ \bibinfo
  {pages} {41294} (\bibinfo {year} {2017})}\BibitemShut {NoStop}%
\bibitem [{\citenamefont {Pitaevskii}(1960)}]{Pitaevskii:1960}%
  \BibitemOpen
  \bibfield  {author} {\bibinfo {author} {\bibfnamefont {L.~P.}\ \bibnamefont
  {Pitaevskii}},\ }\bibfield  {title} {\bibinfo {title} {{Electric Forces in a
  Transparent Dispersive Medium}},\ }\href
  {http://www.jetp.ac.ru/cgi-bin/e/index/e/12/5/p1008?a=list} {\bibfield
  {journal} {\bibinfo  {journal} {JETP}\ }\textbf {\bibinfo {volume} {12}},\
  \bibinfo {pages} {1008} (\bibinfo {year} {1960})}\BibitemShut {NoStop}%
\bibitem [{\citenamefont {van~der Ziel}\ \emph {et~al.}(1965)\citenamefont
  {van~der Ziel}, \citenamefont {Pershan},\ and\ \citenamefont
  {Malmstrom}}]{Malmstrom:1965}%
  \BibitemOpen
  \bibfield  {author} {\bibinfo {author} {\bibfnamefont {J.~P.}\ \bibnamefont
  {van~der Ziel}}, \bibinfo {author} {\bibfnamefont {P.~S.}\ \bibnamefont
  {Pershan}},\ and\ \bibinfo {author} {\bibfnamefont {I.~D.}\ \bibnamefont
  {Malmstrom}},\ }\bibfield  {title} {\bibinfo {title} {{Optically-Induced
  Magnetization Resulting from the Inverse Faraday Effect}},\ }\href
  {https://journals.aps.org/prl/pdf/10.1103/PhysRevLett.15.190} {\bibfield
  {journal} {\bibinfo  {journal} {Phys. Rev. Lett.}\ }\textbf {\bibinfo
  {volume} {15}},\ \bibinfo {pages} {190} (\bibinfo {year} {1965})}\BibitemShut
  {NoStop}%
\bibitem [{\citenamefont {Taguchi}\ and\ \citenamefont
  {Tatara}(2011)}]{MetalsTHzT0:2011}%
  \BibitemOpen
  \bibfield  {author} {\bibinfo {author} {\bibfnamefont {K.}~\bibnamefont
  {Taguchi}}\ and\ \bibinfo {author} {\bibfnamefont {G.}~\bibnamefont
  {Tatara}},\ }\bibfield  {title} {\bibinfo {title} {{Theory of inverse Faraday
  effect in a disordered metal in the terahertz regime}},\ }\href
  {https://journals.aps.org/prb/pdf/10.1103/PhysRevB.84.174433t} {\bibfield
  {journal} {\bibinfo  {journal} {Phys. Rev. B}\ }\textbf {\bibinfo {volume}
  {84}},\ \bibinfo {pages} {174433} (\bibinfo {year} {2011})}\BibitemShut
  {NoStop}%
\bibitem [{\citenamefont {Liang}\ \emph {et~al.}(2021)\citenamefont {Liang},
  \citenamefont {Sukhachov},\ and\ \citenamefont {Balatsky}}]{AME:2021}%
  \BibitemOpen
  \bibfield  {author} {\bibinfo {author} {\bibfnamefont {L.}~\bibnamefont
  {Liang}}, \bibinfo {author} {\bibfnamefont {P.~O.}\ \bibnamefont
  {Sukhachov}},\ and\ \bibinfo {author} {\bibfnamefont {A.~V.}\ \bibnamefont
  {Balatsky}},\ }\bibfield  {title} {\bibinfo {title} {Axial magnetoelectric
  effect in dirac semimetals},\ }\href
  {https://doi.org/10.1103/PhysRevLett.126.247202} {\bibfield  {journal}
  {\bibinfo  {journal} {Phys. Rev. Lett.}\ }\textbf {\bibinfo {volume} {126}},\
  \bibinfo {pages} {247202} (\bibinfo {year} {2021})}\BibitemShut {NoStop}%
\bibitem [{\citenamefont {Juraschek}\ \emph {et~al.}(2017)\citenamefont
  {Juraschek}, \citenamefont {Fechner}, \citenamefont {Balatsky},\ and\
  \citenamefont {Spaldin}}]{PhysRevMaterials.1.014401}%
  \BibitemOpen
  \bibfield  {author} {\bibinfo {author} {\bibfnamefont {D.~M.}\ \bibnamefont
  {Juraschek}}, \bibinfo {author} {\bibfnamefont {M.}~\bibnamefont {Fechner}},
  \bibinfo {author} {\bibfnamefont {A.~V.}\ \bibnamefont {Balatsky}},\ and\
  \bibinfo {author} {\bibfnamefont {N.~A.}\ \bibnamefont {Spaldin}},\
  }\bibfield  {title} {\bibinfo {title} {Dynamical multiferroicity},\ }\href
  {https://doi.org/10.1103/PhysRevMaterials.1.014401} {\bibfield  {journal}
  {\bibinfo  {journal} {Phys. Rev. Mater.}\ }\textbf {\bibinfo {volume} {1}},\
  \bibinfo {pages} {014401} (\bibinfo {year} {2017})}\BibitemShut {NoStop}%
\bibitem [{\citenamefont {Edelstein}(1998)}]{Edelstein:1998}%
  \BibitemOpen
  \bibfield  {author} {\bibinfo {author} {\bibfnamefont {V.~M.}\ \bibnamefont
  {Edelstein}},\ }\bibfield  {title} {\bibinfo {title} {Inverse faraday effect
  in conducting crystals caused by a broken mirror symmetry},\ }\href
  {https://journals.aps.org/prl/pdf/10.1103/PhysRevLett.80.5766} {\bibfield
  {journal} {\bibinfo  {journal} {Phys. Rev. Lett.}\ }\textbf {\bibinfo
  {volume} {80}},\ \bibinfo {pages} {5766} (\bibinfo {year}
  {1998})}\BibitemShut {NoStop}%
\bibitem [{\citenamefont {Qaiumzadeh}\ and\ \citenamefont
  {Titov}(2016)}]{SOsplitZeemantransfer:2016}%
  \BibitemOpen
  \bibfield  {author} {\bibinfo {author} {\bibfnamefont {A.}~\bibnamefont
  {Qaiumzadeh}}\ and\ \bibinfo {author} {\bibfnamefont {M.}~\bibnamefont
  {Titov}},\ }\bibfield  {title} {\bibinfo {title} {Theory of light-induced
  effective magnetic field in rashba ferromagnets},\ }\href
  {https://journals.aps.org/prb/pdf/10.1103/PhysRevB.94.014425} {\bibfield
  {journal} {\bibinfo  {journal} {Phys. Rev. B}\ }\textbf {\bibinfo {volume}
  {94}},\ \bibinfo {pages} {014425} (\bibinfo {year} {2016})}\BibitemShut
  {NoStop}%
\bibitem [{\citenamefont {Pershan}\ \emph {et~al.}(1966)\citenamefont
  {Pershan}, \citenamefont {van~der Ziel},\ and\ \citenamefont
  {Malmstrom}}]{Malmstrom:1966}%
  \BibitemOpen
  \bibfield  {author} {\bibinfo {author} {\bibfnamefont {P.~S.}\ \bibnamefont
  {Pershan}}, \bibinfo {author} {\bibfnamefont {J.~P.}\ \bibnamefont {van~der
  Ziel}},\ and\ \bibinfo {author} {\bibfnamefont {L.~D.}\ \bibnamefont
  {Malmstrom}},\ }\bibfield  {title} {\bibinfo {title} {{Theoretical Discussion
  of the Inverse Faraday Effect, Raman Scattering, and Related Phenomena}},\
  }\href {https://journals.aps.org/pr/pdf/10.1103/PhysRev.143.574} {\bibfield
  {journal} {\bibinfo  {journal} {Phys. Rev.}\ }\textbf {\bibinfo {volume}
  {143}},\ \bibinfo {pages} {574} (\bibinfo {year} {1966})}\BibitemShut
  {NoStop}%
\bibitem [{\citenamefont {Popova}\ \emph {et~al.}(2011)\citenamefont {Popova},
  \citenamefont {Bringer},\ and\ \citenamefont {Bl\"{u}gel}}]{Popova:2011}%
  \BibitemOpen
  \bibfield  {author} {\bibinfo {author} {\bibfnamefont {D.}~\bibnamefont
  {Popova}}, \bibinfo {author} {\bibfnamefont {A.}~\bibnamefont {Bringer}},\
  and\ \bibinfo {author} {\bibfnamefont {S.}~\bibnamefont {Bl\"{u}gel}},\
  }\bibfield  {title} {\bibinfo {title} {{Theory of the inverse Faraday effect
  in view of ultrafast magnetization experiments}},\ }\href
  {https://journals.aps.org/prb/pdf/10.1103/PhysRevB.84.214421} {\bibfield
  {journal} {\bibinfo  {journal} {Phys. Rev. B}\ }\textbf {\bibinfo {volume}
  {84}},\ \bibinfo {pages} {214421} (\bibinfo {year} {2011})}\BibitemShut
  {NoStop}%
\bibitem [{\citenamefont {Popova}\ \emph {et~al.}(2012)\citenamefont {Popova},
  \citenamefont {Bringer},\ and\ \citenamefont {Bl\"{u}gel}}]{Popova:2012}%
  \BibitemOpen
  \bibfield  {author} {\bibinfo {author} {\bibfnamefont {D.}~\bibnamefont
  {Popova}}, \bibinfo {author} {\bibfnamefont {A.}~\bibnamefont {Bringer}},\
  and\ \bibinfo {author} {\bibfnamefont {S.}~\bibnamefont {Bl\"{u}gel}},\
  }\bibfield  {title} {\bibinfo {title} {{Theoretical investigation of the
  inverse Faraday effect via a stimulated Raman scattering process}},\ }\href
  {https://journals.aps.org/prb/pdf/10.1103/PhysRevB.85.094419} {\bibfield
  {journal} {\bibinfo  {journal} {Phys. Rev. B}\ }\textbf {\bibinfo {volume}
  {85}},\ \bibinfo {pages} {094419} (\bibinfo {year} {2012})}\BibitemShut
  {NoStop}%
\bibitem [{\citenamefont {Kirilyuk}\ \emph {et~al.}(2010)\citenamefont
  {Kirilyuk}, \citenamefont {Kimel},\ and\ \citenamefont
  {Rasing}}]{RevModPhys:2010}%
  \BibitemOpen
  \bibfield  {author} {\bibinfo {author} {\bibfnamefont {A.}~\bibnamefont
  {Kirilyuk}}, \bibinfo {author} {\bibfnamefont {A.~V.}\ \bibnamefont
  {Kimel}},\ and\ \bibinfo {author} {\bibfnamefont {T.}~\bibnamefont
  {Rasing}},\ }\bibfield  {title} {\bibinfo {title} {{Ultrafast Optical
  Manipulation of Magnetic Order}},\ }\href
  {https://journals.aps.org/rmp/abstract/10.1103/RevModPhys.82.2731} {\bibfield
   {journal} {\bibinfo  {journal} {Rev. Mod. Phys.}\ }\textbf {\bibinfo
  {volume} {82}},\ \bibinfo {pages} {2731} (\bibinfo {year}
  {2010})}\BibitemShut {NoStop}%
\bibitem [{\citenamefont {Hertel}(2006)}]{Hertel:2006}%
  \BibitemOpen
  \bibfield  {author} {\bibinfo {author} {\bibfnamefont {R.}~\bibnamefont
  {Hertel}},\ }\bibfield  {title} {\bibinfo {title} {{Theory of the Inverse
  Faraday Effect in Metals}},\ }\href
  {https://doi.org/10.1016%2Fj.jmmm.2005.10.225} {\bibfield  {journal}
  {\bibinfo  {journal} {J. Magn. Mag. Mat.}\ }\textbf {\bibinfo {volume}
  {303}},\ \bibinfo {pages} {L1} (\bibinfo {year} {2006})}\BibitemShut
  {NoStop}%
\bibitem [{\citenamefont {Majedi}(2021)}]{Majedi:2021}%
  \BibitemOpen
  \bibfield  {author} {\bibinfo {author} {\bibfnamefont {A.~H.}\ \bibnamefont
  {Majedi}},\ }\bibfield  {title} {\bibinfo {title} {{Microwave-Induced Inverse
  Faraday Effect in Superconductors}},\ }\href
  {https://journals.aps.org/prl/abstract/10.1103/PhysRevLett.127.087001}
  {\bibfield  {journal} {\bibinfo  {journal} {Phys. Rev. Lett.}\ }\textbf
  {\bibinfo {volume} {127}},\ \bibinfo {pages} {087001} (\bibinfo {year}
  {2021})}\BibitemShut {NoStop}%
\bibitem [{\citenamefont {Croitoru}\ \emph
  {et~al.}(2022{\natexlab{b}})\citenamefont {Croitoru}, \citenamefont
  {Lounis},\ and\ \citenamefont {Buzdin}}]{Buzdin:2022}%
  \BibitemOpen
  \bibfield  {author} {\bibinfo {author} {\bibfnamefont {M.~D.}\ \bibnamefont
  {Croitoru}}, \bibinfo {author} {\bibfnamefont {B.}~\bibnamefont {Lounis}},\
  and\ \bibinfo {author} {\bibfnamefont {A.~I.}\ \bibnamefont {Buzdin}},\
  }\bibfield  {title} {\bibinfo {title} {{Influence of a nonuniform thermal
  quench and circular polarized radiation on spontaneous current generation in
  superconducting rings}},\ }\href
  {https://journals.aps.org/prb/abstract/10.1103/PhysRevB.105.L020504}
  {\bibfield  {journal} {\bibinfo  {journal} {Phys. Rev. B.}\ }\textbf
  {\bibinfo {volume} {105}},\ \bibinfo {pages} {L020504} (\bibinfo {year}
  {2022}{\natexlab{b}})}\BibitemShut {NoStop}%
\bibitem [{\citenamefont {Plastovets}\ \emph {et~al.}(2022)\citenamefont
  {Plastovets}, \citenamefont {Tokman}, \citenamefont {Lounis}, \citenamefont
  {Mel'nikov},\ and\ \citenamefont {Buzdin}}]{BuzdinAVortices:2022}%
  \BibitemOpen
  \bibfield  {author} {\bibinfo {author} {\bibfnamefont {V.~D.}\ \bibnamefont
  {Plastovets}}, \bibinfo {author} {\bibfnamefont {I.~D.}\ \bibnamefont
  {Tokman}}, \bibinfo {author} {\bibfnamefont {B.}~\bibnamefont {Lounis}},
  \bibinfo {author} {\bibfnamefont {A.~S.}\ \bibnamefont {Mel'nikov}},\ and\
  \bibinfo {author} {\bibfnamefont {A.~I.}\ \bibnamefont {Buzdin}},\ }\bibfield
   {title} {\bibinfo {title} {{All-optical generation of Abrikosov vortices by
  the inverse Faraday effect}},\ }\href
  {https://journals.aps.org/prb/abstract/10.1103/PhysRevB.106.174504}
  {\bibfield  {journal} {\bibinfo  {journal} {Phys. Rev. B.}\ }\textbf
  {\bibinfo {volume} {106}},\ \bibinfo {pages} {174504} (\bibinfo {year}
  {2022})}\BibitemShut {NoStop}%
\bibitem [{\citenamefont {Yokoyama}(2020)}]{Yokoyama:2018}%
  \BibitemOpen
  \bibfield  {author} {\bibinfo {author} {\bibfnamefont {T.}~\bibnamefont
  {Yokoyama}},\ }\bibfield  {title} {\bibinfo {title} {{Creation of
  Superconducting Vortices by Angular Momentum of Light}},\ }\href
  {https://journals.jps.jp/doi/10.7566/JPSJ.89.103703} {\bibfield  {journal}
  {\bibinfo  {journal} {J. Phys. Soc. Jpn.}\ }\textbf {\bibinfo {volume}
  {89}},\ \bibinfo {pages} {103703} (\bibinfo {year} {2020})}\BibitemShut
  {NoStop}%
\bibitem [{\citenamefont {Banerjee}\ \emph {et~al.}(2022)\citenamefont
  {Banerjee}, \citenamefont {Kumar},\ and\ \citenamefont {Lin}}]{Banerjee2022}%
  \BibitemOpen
  \bibfield  {author} {\bibinfo {author} {\bibfnamefont {S.}~\bibnamefont
  {Banerjee}}, \bibinfo {author} {\bibfnamefont {U.}~\bibnamefont {Kumar}},\
  and\ \bibinfo {author} {\bibfnamefont {S.-Z.}\ \bibnamefont {Lin}},\
  }\bibfield  {title} {\bibinfo {title} {{Inverse Faraday Effect in Mott
  Insulators}},\ }\href
  {https://journals.aps.org/prb/abstract/10.1103/PhysRevB.105.L180414.}
  {\bibfield  {journal} {\bibinfo  {journal} {Phys. Rev. B}\ }\textbf {\bibinfo
  {volume} {105}},\ \bibinfo {pages} {L180414} (\bibinfo {year}
  {2022})}\BibitemShut {NoStop}%
\bibitem [{\citenamefont {Keldysh}(1965)}]{Keldysh:1965}%
  \BibitemOpen
  \bibfield  {author} {\bibinfo {author} {\bibfnamefont {L.}~\bibnamefont
  {Keldysh}},\ }\bibfield  {title} {\bibinfo {title} {{Diagram Technique for
  NonEquilibrium Processes}},\ }\href
  {http://www.jetp.ras.ru/cgi-bin/dn/e_020_04_1018.pdf} {\bibfield  {journal}
  {\bibinfo  {journal} {JETP}\ }\textbf {\bibinfo {volume} {20}},\ \bibinfo
  {pages} {1018} (\bibinfo {year} {1965})}\BibitemShut {NoStop}%
\bibitem [{\citenamefont {Eilenberger}(1968)}]{Eilenberger:1968}%
  \BibitemOpen
  \bibfield  {author} {\bibinfo {author} {\bibfnamefont {G.}~\bibnamefont
  {Eilenberger}},\ }\bibfield  {title} {\bibinfo {title} {{Transformation of
  Gorkov's equation for type II superconductors into transport-like
  equations}},\ }\href {https://link.springer.com/article/10.1007/BF01379803}
  {\bibfield  {journal} {\bibinfo  {journal} {Z. Physik}\ }\textbf {\bibinfo
  {volume} {214}},\ \bibinfo {pages} {195} (\bibinfo {year}
  {1968})}\BibitemShut {NoStop}%
\bibitem [{\citenamefont {Larkin}\ and\ \citenamefont
  {Ovchinnikov}(1969)}]{LarkinOvchinnikov:1969}%
  \BibitemOpen
  \bibfield  {author} {\bibinfo {author} {\bibfnamefont {A.~I.}\ \bibnamefont
  {Larkin}}\ and\ \bibinfo {author} {\bibfnamefont {Y.~N.}\ \bibnamefont
  {Ovchinnikov}},\ }\bibfield  {title} {\bibinfo {title} {{Quasiclassical
  Method in the Theory of Superconductivity}},\ }\href
  {https://link.springer.com/article/10.1007/BF01379803} {\bibfield  {journal}
  {\bibinfo  {journal} {JETP}\ }\textbf {\bibinfo {volume} {28}},\ \bibinfo
  {pages} {1200} (\bibinfo {year} {1969})}\BibitemShut {NoStop}%
\bibitem [{\citenamefont {Yoshino}(2011)}]{YOSHINO20112531}%
  \BibitemOpen
  \bibfield  {author} {\bibinfo {author} {\bibfnamefont {T.}~\bibnamefont
  {Yoshino}},\ }\bibfield  {title} {\bibinfo {title} {Simple theory of the
  inverse faraday effect with relationship to optical constants
  \begin{math}n\end{math} and \begin{math}k\end{math}},\ }\href
  {https://doi.org/https://doi.org/10.1016/j.jmmm.2011.05.010} {\bibfield
  {journal} {\bibinfo  {journal} {Journal of Magnetism and Magnetic Materials}\
  }\textbf {\bibinfo {volume} {323}},\ \bibinfo {pages} {2531} (\bibinfo {year}
  {2011})}\BibitemShut {NoStop}%
\bibitem [{\citenamefont {Basini}\ \emph {et~al.}(2022)\citenamefont {Basini},
  \citenamefont {Pancaldi}, \citenamefont {Wehinger}, \citenamefont {Udina},
  \citenamefont {Tadano}, \citenamefont {Hoffmann}, \citenamefont {Balatsky},\
  and\ \citenamefont {Bonetti}}]{Bonetti:2022}%
  \BibitemOpen
  \bibfield  {author} {\bibinfo {author} {\bibfnamefont {M.}~\bibnamefont
  {Basini}}, \bibinfo {author} {\bibfnamefont {M.}~\bibnamefont {Pancaldi}},
  \bibinfo {author} {\bibfnamefont {B.}~\bibnamefont {Wehinger}}, \bibinfo
  {author} {\bibfnamefont {M.}~\bibnamefont {Udina}}, \bibinfo {author}
  {\bibfnamefont {T.}~\bibnamefont {Tadano}}, \bibinfo {author} {\bibfnamefont
  {M.~C.}\ \bibnamefont {Hoffmann}}, \bibinfo {author} {\bibfnamefont {A.~V.}\
  \bibnamefont {Balatsky}},\ and\ \bibinfo {author} {\bibfnamefont
  {S.}~\bibnamefont {Bonetti}},\ }\href {https://arxiv.org/pdf/2210.01690.pdf}
  {\bibinfo {title} {{Terahertz electric-field driven dynamical multiferroicity
  in $SrTiO_3$}}} (\bibinfo {year} {2022})\BibitemShut {NoStop}%
\bibitem [{\citenamefont {Kozina}\ \emph {et~al.}(2019)\citenamefont {Kozina},
  \citenamefont {Fechner}, \citenamefont {Marsik}, \citenamefont {van Driel},
  \citenamefont {Glownia}, \citenamefont {Bernhard}, \citenamefont {Radovic},
  \citenamefont {Zhu}, \citenamefont {Bonetti}, \citenamefont {Staub},\ and\
  \citenamefont {Hoffmann}}]{BonettiNatPhys:2019}%
  \BibitemOpen
  \bibfield  {author} {\bibinfo {author} {\bibfnamefont {M.}~\bibnamefont
  {Kozina}}, \bibinfo {author} {\bibfnamefont {M.}~\bibnamefont {Fechner}},
  \bibinfo {author} {\bibfnamefont {P.}~\bibnamefont {Marsik}}, \bibinfo
  {author} {\bibfnamefont {T.}~\bibnamefont {van Driel}}, \bibinfo {author}
  {\bibfnamefont {J.~M.}\ \bibnamefont {Glownia}}, \bibinfo {author}
  {\bibfnamefont {C.}~\bibnamefont {Bernhard}}, \bibinfo {author}
  {\bibfnamefont {M.}~\bibnamefont {Radovic}}, \bibinfo {author} {\bibfnamefont
  {D.}~\bibnamefont {Zhu}}, \bibinfo {author} {\bibfnamefont {S.}~\bibnamefont
  {Bonetti}}, \bibinfo {author} {\bibfnamefont {U.}~\bibnamefont {Staub}},\
  and\ \bibinfo {author} {\bibfnamefont {M.~C.}\ \bibnamefont {Hoffmann}},\
  }\bibfield  {title} {\bibinfo {title} {{Terahertz-driven phonon upconversion
  in $SrTiO_3$}},\ }\href {https://www.nature.com/articles/s41567-018-0408-1}
  {\bibfield  {journal} {\bibinfo  {journal} {Nat. Phys.}\ }\textbf {\bibinfo
  {volume} {15}},\ \bibinfo {pages} {387} (\bibinfo {year} {2019})}\BibitemShut
  {NoStop}%
\bibitem [{\citenamefont {Serene}\ and\ \citenamefont
  {Rainer}(1983)}]{RainerSerene:1983}%
  \BibitemOpen
  \bibfield  {author} {\bibinfo {author} {\bibfnamefont {J.~W.}\ \bibnamefont
  {Serene}}\ and\ \bibinfo {author} {\bibfnamefont {D.}~\bibnamefont
  {Rainer}},\ }\bibfield  {title} {\bibinfo {title} {The quasiclassical
  approach to superfluid \begin{math}{^3He}\end{math}},\ }\href@noop {}
  {\bibfield  {journal} {\bibinfo  {journal} {Phys. Reps.}\ }\textbf {\bibinfo
  {volume} {101}},\ \bibinfo {pages} {221} (\bibinfo {year}
  {1983})}\BibitemShut {NoStop}%
\end{thebibliography}%
\end{document}